\documentclass[11pt,twoside]{article} 
\usepackage{acta-info}
\usepackage{euler}
\input{psfig.sty}
\usepackage {graphicx}
\usepackage{amsmath} 
\usepackage{hyperref}
\usepackage{amssymb}

\newcommand{\vol}{\textbf{9,} 2 (2017) 134--143}   %% to be completed by the editor, do not modify it
\newcommand{\amss}{\amssubj{%
68Q05, 68W10
}}     %% Source:   http://www.ams.org/msc/
\newcommand{\CRs}{\CRsubj{%
F.1.1
}}   %% Source: http://www.acm.org/about/class/1998
\newcommand{\keyw}{\keywords{%
optical computing, set splitting, NP-complete
}} 

\begin{document}

\title{An optical solution for the set splitting problem}

%\author{\\
%Faculty of Exact Sciences and Engineering,\\
%,\\
%Alba-Iulia, Romania.\\
%\\
%\url{}\\
%}

\maketitle

\oneauthor{%
 %% Write here the name of the author and his homepage address using \href command  
\href{https://mihaioltean.github.io/optical}{Mihai OLTEAN} 
}{%
%% Write here your affiliation and its homepage address including maybe your address using \href command. 
%% You can use \\ for line breaks.
\href{http://www.uab.ro}{1 Decembrie 1918 University of Alba Iulia\\ Alba Iulia, Romania}
}{%
 %% Write here your email.
 \href{mailto:mihai.oltean@gmail.com}{mihai.oltean@gmail.com}
 \href{https://mihaioltean.github.io/optical}{https://mihaioltean.github.io/optical} 
}

\short{%
%% Write here the short name of authors, using commas.
M. Oltean
}{%
%%Write here the short title
An optical solution for the set splitting problem
}

\begin{abstract}

We describe here an optical device, based on time-delays, for solving the set splitting problem which is well-known NP-complete problem. The device has a graph-like structure and the light is traversing it from a start node to a destination node. All possible (potential) paths in the graph are generated and at the destination we will check which one satisfies completely the problem's constrains.

\end{abstract}

\section{Introduction}

Recently, an increased number of difficult (most of them being NP-complete \cite{garey}) problems have been proposed to be solved by using optical devices. Hamiltonian path \cite{dolev,oltean_hamiltonian,shaked, shakeri}, traveling salesman \cite{dolev,haist,haist_erratum}, subset sum \cite{dolev,raqibul,muntean_unbounded,oltean_subset_sum,muntean_thesis}, exact cover \cite{muntean_thesis,oltean_exact_cover}, Diophantine equations \cite{muntean_diophantine,muntean_thesis}, 3-SAT \cite{dolev,sama_3_sat,sama_3_sat_2,head_3_sat}, SAT \cite{sama_bounds, sama_satisfiability, sama_complexity, oltean_sat}, dominating set \cite{sama_dominating}, prime factorization \cite{nitta1,nitta2}, security \cite{haist_security}, independent sets \cite{head_independent}, graph colorability \cite{sama_colorability}, k-clique \cite{sama_sheets}, ultrafast arithmetic \cite{haist_arithmetic}, abstract machines \cite{sama_reduce} are just few of the problems whose solution can be computed by using an optical device.

Here we show how to solve another problem, namely set splitting, which is an known NP-complete problem \cite{garey}. The underlying mechanism is to use delays for encoding possible solutions. The approach here is derived from the solution for the subset sum problem \cite{oltean_subset_sum}. The novelty consists in a special set of delays attached to each arc and a special set of moments when the solution is expected in the destination node.

The paper is organized as follows:  Section \ref{set_splitting_problem} contains a description of the problem to be solved. Properties of the signal useful for our research and the operations performed on that signal are described in section \ref{properties}. Section \ref{proposed} deeply describe the proposed device. Details about the physical implementation are given in section \ref{hard}. Complexity is discussed in section \ref{complexity}. The size of the instances that can be solved with this method is computed in section \ref{psize}. A short discussion of the weaknesses of this method is given in section \ref{weak}. Section \ref{conclusion} concludes our paper.

\section{The set splitting problem}
\label{set_splitting_problem}

"Given a family $F$ of subsets of a finite set $A$, decide whether there exists a partition of $A$ into two subsets $A_1$, $A_2$ such that all elements of $F$ are split by this partition, i.e., none of the elements of $F$ is completely in $A_1$ or $A_2$." \cite{garey,set_splitting_wikipedia}.

set splitting is a NP-complete problem. No polynomial-time algorithm is known to exist for it.

The optical solution of the set splitting is based on the solution for the subset sum problem whose definition is given below:

Given a set of positive numbers $A = \{a_1, a_2, ..., a_n\}$ and another positive number $B$. Is there a subset of $A$ whose sum equals $B$? \cite{garey}

\section{Time-delay systems}
\label{properties}

Time-delay systems have been proposed in \cite{oltean_hamiltonian}. They are based on 2 properties of the signals (optical, electrical, etc):
\begin{itemize}\addtolength{\itemsep}{-0.6\baselineskip}

\item{The signal can be delayed by forcing it to pass through a cable of a certain length.}

\item{The signal can be easily divided into multiple signals of smaller intensity/power which run within the same cables. This ensures a high parallelism of the method.}

\end{itemize}

The proposed device has a graph-like structure with a start and destination node. Nodes are connected through cables made of optical fiber.

The system works as follows: A beam of light is sent to the start node. The beams travel through arcs and is divided inside nodes. Because arcs have a length greater than 0, the light is delayed by them. At the destination node we will have different signals arriving at various moments of time. One of them will tell us if we have a solution to our problem.

Since we work with continuous signal we cannot expect to have discrete output at the destination node. This means that beams arrival is notified by fluctuations in the intensity of the light. These fluctuations will be transformed, by a photodiode, in fluctuations of the electric power which will be read by an oscilloscope.

\section{The optical device for the set splitting problem}
\label{proposed}

This section deeply describes the proposed system.

Since the solution for set splitting is based on the solution from the subset sum, we first briefly describe (see section \ref{graph_subset_sum}) the device which solves the subset sum problem. This device has been proposed in \cite{oltean_subset_sum}. Later, in section \ref{graph_set_splitting} we describe the graph for the set splitting problem.

\subsection{The graph for the subset sum problem}
\label{graph_subset_sum}

As described in \cite{oltean_subset_sum} the numbers from the given set $A$ represent the delays induced to the signals (light) that passes through the device. For instance, if numbers $a_1$, $a_3$ and $a_7$ generate the expected subset, then the total delay of the signal should be $a_1 + a_3 + a_7$. Recall that, when using light, we can easily induce some delays by forcing the beam to pass through an optical cable of a given length. 

The device has been designed \cite{oltean_subset_sum} as a simple directed graph. There are $n + 1$ nodes connected by arcs. In each node (but the last one) we place a beam-splitter which will split a beam into 2 sub-beams of smaller intensity. Arcs are implemented by using optical cables and have lengths proportional to the numbers from the given set $A$. A mechanism for skipping a value from the given set is also needed. A possible way for achieving this is to add an extra arc, of length 0, between any pair of consecutive nodes. A beam leaving a node will have the possibility to either traverse an arc representing a number from the given set or to skip it (by traversing the arc of 0 length).

\begin{figure}[htbp]
\centerline{\includegraphics[width=4.52in,height=1.21in]{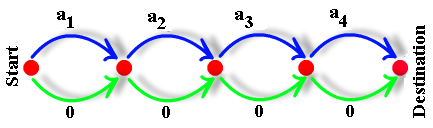}}
\caption{The device for solving the subset sum problem. Each arc delays the beam by the amount of time written on it. Image taken from \cite{oltean_subset_sum}.}
\label{fig_subset_sum_device}
\end{figure}

The graph for solving the subset sum problem is depicted in Figure \ref{fig_subset_sum_device}. 

In the destination node we will have signals (fluctuation in the intensity of the signal) arriving at moments representing the sum of elements of all subsets. For instance if the lights travels through the top arcs only we will have the subset containing all elements: $a_1, a_2, ..., a_n$. If the light travels through bottom arc only we will have the empty subset. Thus the device will generate all possible subsets of $A$. Each subset will delay one of the beams by an amount of time equal to the sum of the lengths of the arcs in that path. If there is a fluctuation in the intensity of the signal at moment $B$ it means that we have a solution to our problem.

Note that in practice we cannot have cables of length 0 in practice, thus we add a small $\epsilon$ to the length of all arcs and the final solution is expected at moment $B + n * \epsilon$.

\subsection{The graph for the set splitting problem}
\label{graph_set_splitting}

We have seen that the device described in section \ref{graph_subset_sum} generates all possible subsets of the given set $A$, so theoretically we could use it for the set splitting problem. But, there are some issues:

\begin{itemize}\addtolength{\itemsep}{-0.6\baselineskip}

\item{In the case of subset sum we already have some values for the elements in set A. Here, in the set splitting, we have no predefined values in $A$.}
\item{If 2 values from the set $A$ are identical, it is not possible to uniquely identify a subset at the destination node.}

\end{itemize}

In order to overcome these issues we introduce here a new set of delays. The main attribute of this delaying system is to allow a unique identification of the generated subsets.

The smallest set of $n$ numbers such that each subset has a different value for the sum of its elements is the set of the first $n$ powers of 2: $A=\{2^0, 2^1, \ldots 2^{n-1}\}$.

Thus, we assign the values \{$2^0, 2^1, \ldots 2^{n-1}$\} as lengths for the top arcs.

\begin{figure}[htbp]
\centerline{\includegraphics[width=4.52in,height=1.21in]{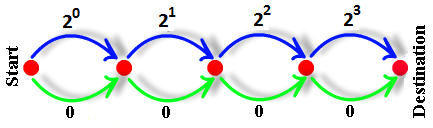}}
\caption{The device for solving the set splitting problem. Each arc delays the beam by the amount of time written on it.}
\label{fig_set_splitting_device}
\end{figure}

The graph for solving the set splitting problem is depicted in Figure \ref{fig_set_splitting_device}. 

Now we can easily identify a subset at the destination node: having the moment of arrival (name it $k$) we know exactly what elements belong to that subset (because each number ($k$) is uniquely decomposed as sum of powers of 2).

Knowing the subset, we can easily find if it represents or not a solution to our problem. Let's denote by $A_1$ the subset that has just arrived in the destination node. We also compute its complement: $A_2 = A - A_1$.

Now, we want to know if the split of $A$ into sets ($A_1$ and $A_2$) represents or not a solution to our problem, that is, whether one of the sets in $F$ is completely included in either $A_1$ or $A_2$. For this we check if any of them ($A_1$ or $A_2$) includes any set from $F$. If they do, it means that the corresponding split does not represent a solution for our problem.

We can check the inclusion property very easily by working only with delay moments: for each set in $F$ we compute and cache the moments when any superset (a set that includes the current one) of it arrives at the destination node. Let's denote by $M$ those moments. If a signal arrives at a moment not belonging to $M$, that set (who generated the signal) is a solution for our problem. If $M$ contains all integer numbers between 0 and $2^n-1$ it means that the instance does not have a solution.

We can do a small optimization here: we want to minimize the number of moments when we wait for a solution at the destination node. So, we check the size of $M$. If it is less than $2^{n-1}$ we wait for numbers belonging to $M$. Otherwise we wait for numbers not belonging to $M$.

\subsection{Example of solution}

Let's suppose that we have a set $A$ made from 4 elements: $A = \{a_1, a_2, a_3, a_4\}$ and a set of subsets $F = \{F_1 = \{a_1, a_2\}, F_2 = \{a_1, a_3\}\}$. We want to find if it is possible to split $A$ into 2 subsets ($A_1$ and $A_2$) such that neither $F_1$ nor $F_2$ is completely included in $A_1$ or $A_2$.

To each number from $A$ we assign a power of 2 as follows: $a_1 = 2^0, a_2 = 2^1, a_3 = 2^2$ and $a_4 = 2^3$.

The supersets which includes $F_1$ are: 

\centerline{$\{\{a_1, a_2\}, \{a_1, a_2, a_3\}, \{a_1, a_2, a_4\}, \{a_1, a_2, a_3, a_4\}\} $}

\noindent which have the following delay times: $M_1 = \{3, 7, 11, 15\}$ (which were computed as sum of elements).

The supersets including $F_2$ are: 

\centerline{$\{\{a_1, a_3\}, \{a_1, a_3, a_2\}, \{a_1, a_3, a_4\}, \{a_1, a_3, a_2, a_4\}\} $}
\noindent  which have the following delay times: $M_2 = \{5, 7, 13, 15\}$.

Set $M$ which represents the union of $M_1$ with $M_2$ is $M = \{3, 5, 7, 11, 13, 15\}$.

If a beam of light arrives in the destination at any of the moments from $M$ it does not encode a subset representing a solution to our problem. However, if a beam arrives at another moment in the integer interval [1...15] but does not belong to $M$, it represents a solution to our problem. For instance, if a beam arrives at moment 1 it represents a solution because it encodes the split $A_1 = \{a_1\}$ and $A_2 = \{a_2, a_3, a_4\}$ and none of the sets in $F$ is included into either $A_1$ or $A_2$.

\section{Physical implementation}
\label{hard}

As discussed in \cite{oltean_subset_sum} for the physical implementation we need the following components:

\begin{itemize}\addtolength{\itemsep}{-0.6\baselineskip}

\item{A source of light (laser),}

\item{Several beam-splitters for dividing light beams into 2 sub-beams.}

\item{A high speed photodiode for converting light into electrical power. The photodiode is placed in the destination node,}

\item{A tool for detecting fluctuations in the intensity of electric power generated by the photodiode (oscilloscope),}

\item{A set of optical fiber cables having lengths proportional with the numbers $2^0, 2^1, \ldots 2^{n-1}$ (plus constant $\epsilon$) and another set of $n$ cables having fixed length $\epsilon$. }

\end{itemize}

\section{Build and running complexity}
\label{complexity}

The time required to build the device has $\theta(n*2^n)$ complexity because we have to build $n$ arcs having at most $2^{n-1}$ length.

The running time complexity is $\theta(2^n)$.

The intensity of the signal decreases exponentially with the number of nodes (because it is divided in 2 in each node). This is why the required power is proportional to $2^n$ (exponential).

\section{Instance size}
\label{psize}

We want to find the size of the instances (how many numbers can we have in set $A$) that can be solved by our device.

For computing this number we use the same reasoning as for the Hamiltonian Path problem \cite{sama_hamiltonian} because in both cases the delays are powers of 2. (Note that only delays are similar with Hamiltonian Path solution. The devices (graphs) are extremely different.)

Assuming that the best oscilloscope available on the market has the rise-time in the range of picoseconds ($10^{-12}$ seconds - this is also the smallest difference between 2 consecutive moments when 2 solution may arrive in the destination) and also knowing the speed of light ($3 \cdot 10^{8} m/s$) we can compute the minimal cable length that should be traversed by the light in order to be delayed by  $10^{-12}$ seconds. This is 0.0003 meters \cite{oltean_hamiltonian}.

This length is for the shortest cable. All other cables have lengths obtained by multiplying this shortest length with powers of 2.

It is easy to compute the maximal number of nodes that a graph can have in order to have the total delay equal by 1 second. This results from the equation (taken from \cite{oltean_hamiltonian}):
\[
2^n \cdot 0.0003 = 3 \cdot 10^8
\]
This gives us a number of nodes equal to 39 nodes, so in one second we can solve instances having 39 numbers. However, the length of the optical fibers used for inducing the largest delay for this graph is huge: about $8 \cdot 10^{8}$ meters. We cannot expect to have such long cables for our experiments \cite{oltean_hamiltonian}.

But, shorter cables (of several hundreds of kilometers) are already available in the internet networks. They can be easily used for our purpose. Assuming that the longest cable that we have is about 300 kilometers we may solve instances with about 26 numbers. The amount of time required to obtain a solution is about $10^{-6}$ seconds \cite{oltean_hamiltonian}.

The maximum number of nodes can be increased by increasing the precision of the instruments (oscilloscope, photodiode etc).

In \cite{braich} the authors have solved with DNA another problem with a $2^n$ complexity, namely the SAT problem. The largest instance solved was with 20 nodes. Here we have shown that theoretically we can solve an instance with 26 nodes with the available resources. This is with almost 2 orders of magnitude larger that the DNA solution presented in \cite{braich} for the SAT problem.

\section{Weaknesses of the solution}
\label{weak}

The proposed device has some weaknesses:

\begin{itemize}\addtolength{\itemsep}{-0.6\baselineskip}

\item{it contains exponential delays,}

\item{the number of possible moments when a solution can appear is exponential in the number of elements in the initial set. This is different compared to subset sum (see \cite{oltean_subset_sum}) where only one moment was enough for detecting the solution.}

\end{itemize}

\section{Conclusion}
\label{conclusion}

We have shown how the set splitting problem can be attacked with an optical device. The solution was derived from the graph for the the subset sum problem by adding a new set of delays and a new set of moments when the solution is expected at the destination node. Weak points of the proposed solution are: the use of optical cables of exponential length, and the exponential amount of energy required to power the device.

\bigskip
\rightline{\emph{Received: November 7, 2017 {\tiny \raisebox{2pt}{$\bullet$\!}} Revised: }} %% to be completed by the editor 

\end{document}